\documentclass[
  aip,
  amsmath,amssymb,
  reprint,
]{revtex4-1}

\usepackage{graphicx}
\usepackage{dcolumn}
\usepackage{bm}

\usepackage[utf8]{inputenc}
\usepackage[T1]{fontenc}
\usepackage{amsmath, mathtools, amssymb}
\usepackage{etoolbox}

\makeatletter
\def\@email#1#2{%
  \endgroup
  \patchcmd{\titleblock@produce}
  {\frontmatter@RRAPformat}
  {\frontmatter@RRAPformat{\produce@RRAP{*#1\href{mailto:#2}{#2}}}\frontmatter@RRAPformat}
}
\makeatother
\begin{document}

\preprint{AIP/123-QED}

\title[Spatial Sound Modulation through Manual Reconfiguration with Phased Plate]{Spatial Sound Modulation through Manual Reconfiguration of Phased Plate}
\author{Soma Mochizuki}
\affiliation{School of Informatics, College of Media Arts, Science and Technology, University of Tsukuba, Kasuga Campus Kasuga 1-2, Tsukuba, 305-8550, Ibaraki, Japan}
\author{Yoichi Ochiai}
\affiliation{Institute of Library, Information and Media Science, University of Tsukuba, Kasuga Campus Kasuga 1-2, Tsukuba, 305-8550, Ibaraki, Japan}
\affiliation{R\&D Center for Digital Nature, University of Tsukuba, Kasuga Campus Kasuga 1-2, Tsukuba, 305-8550, Ibaraki, Japan}
\affiliation{Pixie Dust Technologies, Inc, Tokyo, 104-0028, Japan}
\author{Tatsuki Fushimi}
\affiliation{Institute of Library, Information and Media Science, University of Tsukuba, Kasuga Campus Kasuga 1-2, Tsukuba, 305-8550, Ibaraki, Japan}
\affiliation{R\&D Center for Digital Nature, University of Tsukuba, Kasuga Campus Kasuga 1-2, Tsukuba, 305-8550, Ibaraki, Japan}
\email{tfushimi@slis.tsukuba.ac.jp}

\date{\today}

\begin{abstract}
Ultrasonic phased array technology, while versatile, often requires complex computing resources and numerous amplifier components. We present a Manually Reconfigurable Phased Array that physically controls transducer position and phase, offering a simpler alternative to traditional phased array transducers (PAT). Our system uses a conductor rod-connected transducer array with an underlying plate that modulates the phase state through its shape and electrode arrangement. This approach enables variable phase reconstruction with reduced computational demands and lower cost. Experimental results demonstrate the device's capability to focus ultrasonic waves at different spatial locations. The system's design facilitates the creation of acoustic fields without extensive digital control, potentially broadening applications in areas such as aerial haptics, audio spotlighting, and educational demonstrations of acoustic phenomena. This work contributes to the development of more accessible and computationally efficient acoustic phased array systems.
\end{abstract}

\maketitle
\begin{figure}[b]
\includegraphics[width=0.45\textwidth]{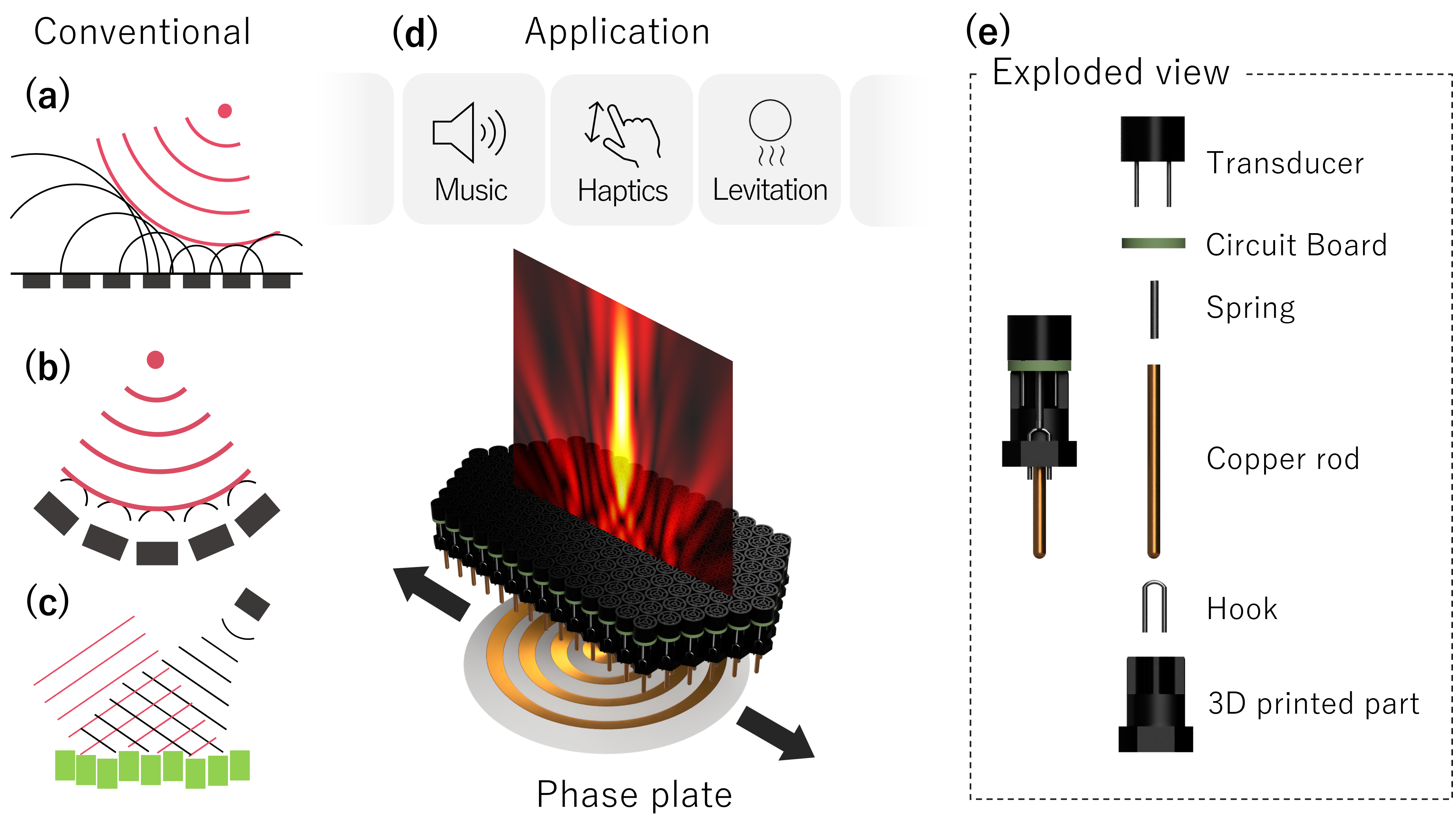}
\caption{Overview of Experimental Setup. (a) Conventional PAT configuration with electronically steerable array (b) Spatially configured array (c) Reflective approach (d) Overview of proposed method (e) Exploded view of the transducer modulation element}
\label{fig_exp_overview}    
\end{figure}

The field of holographic representations of acoustic fields has experienced a significant surge in interest, driven by advancements in ultrasonic tactile displays~\cite{Hoshi2010, Long2014}, acoustic levitation~\cite{Marzo2015, Inoue2019}, digital microfluidics~\cite{Koroyasu2023}, and volumetric display technologies~\cite{Ochiai2014a,Fushimi2019a, Hirayama2019}. Traditionally, mid-air acoustic holograms are encoded using ultrasonic phased array transducers (PATs), known for their dynamic systems with rapid update capabilities and precise phase representation, as depicted in Fig.\ref{fig_exp_overview}a\cite{Marzo2018, Suzuki2021, Montano-Murillo2023}. However, these systems face a challenge in controlling multiple transducers through FPGA boards. Alternative approaches, such as the spatially configured transducer array exemplified by TinyLev (Fig.\ref{fig_exp_overview}b), offer affordability and simplicity in implementing acoustic levitation but are constrained in their dynamic capabilities\cite{Marzo2017}.

Recent advancements in spatial sound modulators (SSMs) have shaped acoustic field modulation methods, as shown in Fig.~\ref{fig_exp_overview}c. Prat-Camps et al. introduced a manually reconfigurable reflective SSM for standing wave-type fields \cite{PratCamps2020}. Hardwick et al~developed cost-effective, segmented SSMs, but their reflective nature limited field diversity \cite{Hardwick2023}. Choi et al.~explored dynamic labyrinthine transmissive ultrasonic metamaterials, enhancing SSM reconfigurability, though challenges in actuating central elements persisted \cite{Choi2024}. These developments have advanced SSM technology while highlighting ongoing issues in field diversity and full dynamic manipulation.
\begin{figure}[t]
\includegraphics[width=0.9\linewidth]{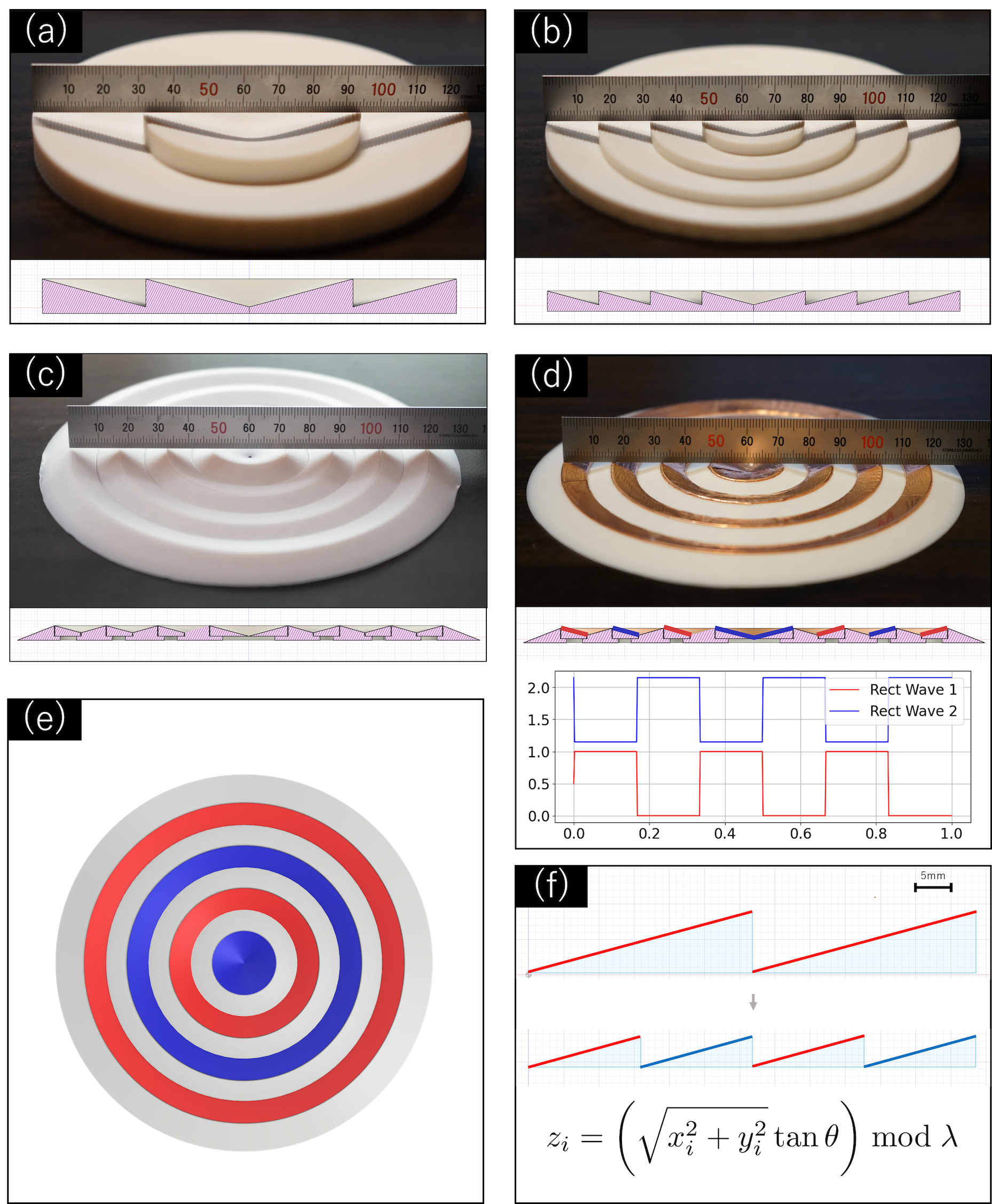}
\caption{Phase plate design and implementation. (a) Initial thick phase plate design. (b) Thinned phase plate design using $\lambda/2$ wavelength assumption. (c) Phase plate after $15.7^\circ$ slope smoothing, optimized through trial and error for movement under the transducer array. (d) Final phase plate with copper foil applied to conductive parts. (e) Illustration of signal phase distribution: red and blue areas represent opposite phases, white areas are insulated. (f) Mechanism of plate thinning: red surface receives original signal, blue surface receives $\pi$-shifted signal. }
\label{fig:phasePlate}
\end{figure}

Here, diverging from these reflective approaches, our work introduces a manually reconfigurable phased array that focuses on dynamic reconfiguration capabilities at the transducer side, as illustrated in Fig.~\ref{fig_exp_overview}d. This system encodes the acoustic hologram's phase through the spatial height of the transducer array. By employing retractable transducers mounted on spring mechanisms, our design enables these transducers to align their height with the phase plate positioned beneath them. This method not only allows for a broader range of acoustic field manipulations, moving beyond the limitations of standing wave-type fields typically associated with reflective systems.
\begin{figure*}[t]
\includegraphics[width=0.8\textwidth]{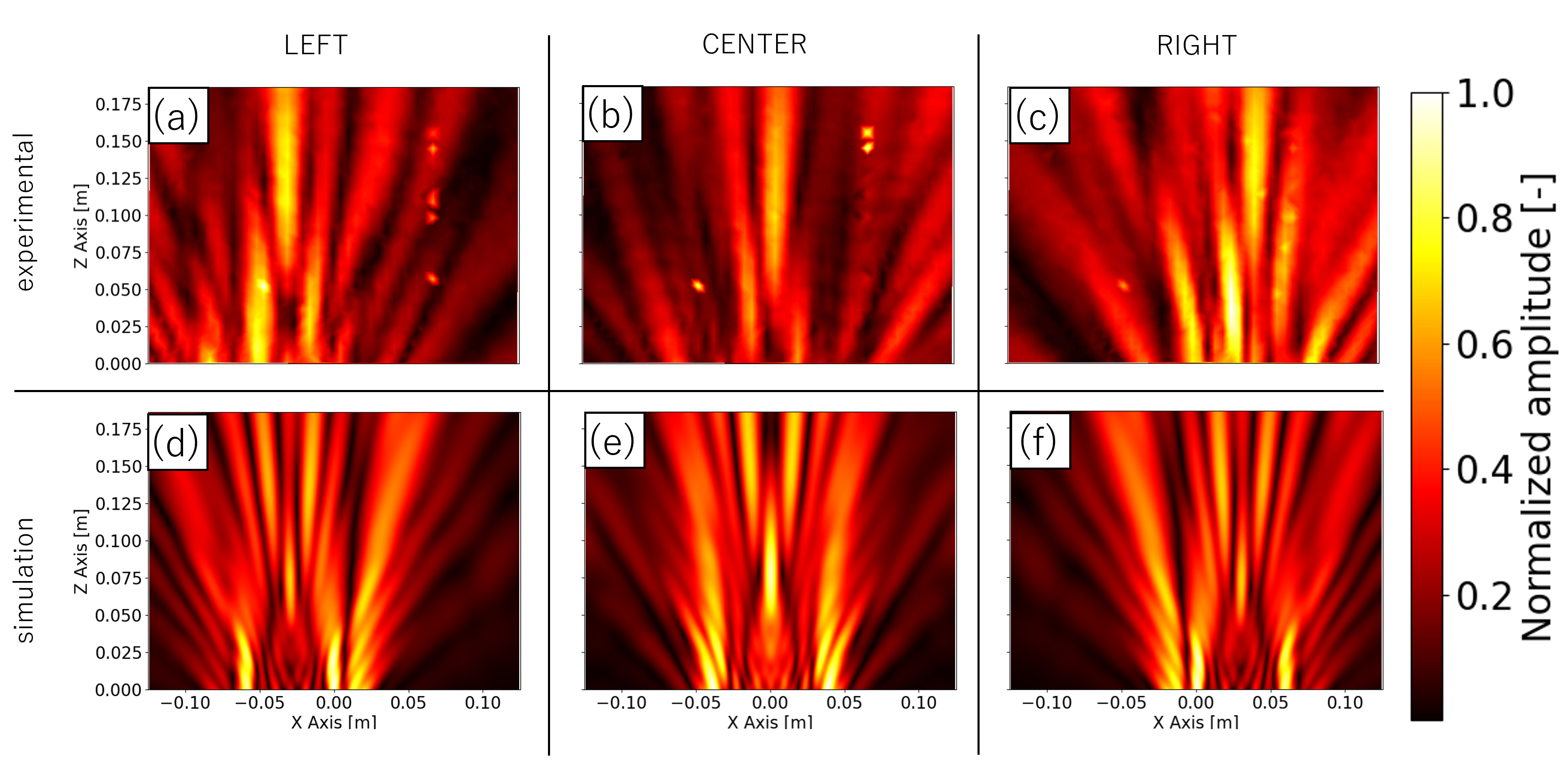}
\caption{Comparison of experimental and simulated acoustic field distributions for different focus positions. Top row (a-c): Experimental results measured using Laser Doppler Vibrometer (LDV). Bottom row (d-f): Corresponding simulated results. Columns represent different focus positions: (left) focus shifted left, (center) focus at center, (right) focus shifted right. The overall agreement between experimental and simulated results validates the efficacy of our manually reconfigurable phased array system.}
\label{LDV}
\end{figure*}

The manually reconfigurable phased array consists of multiple spring-loaded transducers as detailed in the exploded view shown in Fig.~\ref{fig_exp_overview}e. Assembly instructions and further component details are available in the supplementary material. Each unit within the array features a 40 kHz transducer (UT1007-Z325R), encased in a 3D printed housing, and is supported by a spring, a copper rod, a PCB board, and a hook. Both the hook and the spring are connected to a common ground for all transducers, ensuring stability and uniformity. The elongated copper rod is maneuvered up and down by the phase plate positioned below the array. This setup allows the transducers to be in phase but also confines their activation to the specific area influenced by the phase plate, thereby enhancing the array's effectiveness and the quality of the acoustic field produced.

Encoding the acoustic hologram requires the understanding that the phase plate can only be pushed up or down. This limits the direct conversion from acoustic hologram to phase plates (i.e.~transducer cannot be shifted radially easily), and we calculated the required shifting for a single focus: 
\begin{equation}
z_{i}=z_{f}+\sqrt{(d_{i}-d_{i}\bmod\lambda)^2-(x_i-x_f)^2-(y_i-y_f)^2}.
\end{equation}
Here, $(x_{i},y_{i},z_{i})$ represent the transducer's coordinates, $(x_{f},y_{f},z_{f})$ denote the focus point's coordinates, $d=(x_i-x_p)^2+(y_i-y_p)^2+(z_i-z_p)^2$ is the distance between the focus point and the transducer, and $\lambda$ symbolizes the acoustic wave's wavelength. Similarly, the phase plate for bessel beam is calculated using: 

\begin{equation}\label{eq:bessel}
z_{i}=\left(\sqrt{x_{i}^2+y_{i}^2}\tan\theta\right)\bmod\lambda
\end{equation}
$\theta$ is the cone angle of the bessel beam.
The bessel phase plate was made of Poly Lactic Acid (PLA) using a Fused Deposition Modeling (FDM) 3D printer as shown in Figure~\ref{fig:phasePlate}a. This bessel phase plate's shape is the result of culculation with $\theta = \frac{\pi}{12}$ and $\lambda = 8.5~\mbox{mm}$ in the equation~\ref{eq:bessel}.

Our manually reconfigurable phased array employs a signal transmission method that differs from traditional FPGA-controlled arrays. Each transducer, connected to a probe (dynamic rod), receives signals only when in contact with the phase plate's conductive surfaces, preventing unintended activation and minimizing destructive interference.

We developed an approach to thin down the phase plate while maintaining the desired acoustic effect. Fig.~\ref{fig:phasePlate}a shows the initial, thicker design of the phase plate. By increasing the number of signal levels to $N$, we can effectively reduce the thickness of the phase plate by a factor of $N$ while preserving the same acoustic effect. For instance, when $N = 2$, the signal levels are split into two distinct phases (0 and $\pi$), which allows us to create a phase plate that is half as thick as the original (Fig.~\ref{fig:phasePlate}a), while still producing the same acoustic modulation, as shown in Fig.~\ref{fig:phasePlate}b.
To facilitate movement beneath the transducer array, we needed to smooth the phase plate's shape. This was achieved by shaving off the corners of the calculated shape in CAD. Importantly, the optimal angle for this smoothing was determined through extensive trial and error. We found that a slope angle of 15.7 degrees worked best for our setup, but it's crucial to note that this angle can vary depending on factors such as the spring stiffness of the transducer array and the shape of the probe tip. Fig.~\ref{fig:phasePlate}c illustrates the result of this smoothing process, showing the phase plate with its 15.7-degree slopes.

The implementation required creating a plate with both conductive and non-conductive surfaces. The final phase plate, depicted in Fig.\ref{fig:phasePlate}d, was created by applying copper foil to the conductive parts. Fig.\ref{fig:phasePlate}e illustrates the phase of the signal flowing through these conductive parts, with red and blue areas representing opposite phases, and white areas indicating insulated surfaces.

The mechanism behind this thinning process is further explained in Fig.~\ref{fig:phasePlate}f. It shows two isolated conductive surfaces: one receiving the original signal (red) and another receiving a signal with a phase shift of $\pi$ (blue). This design allows us to achieve the desired acoustic field with a thinner plate by adjusting the electronic signals.

\begin{figure}[hpb]
  \begin{minipage}[b]{0.8\columnwidth}
    \includegraphics[width=\columnwidth]{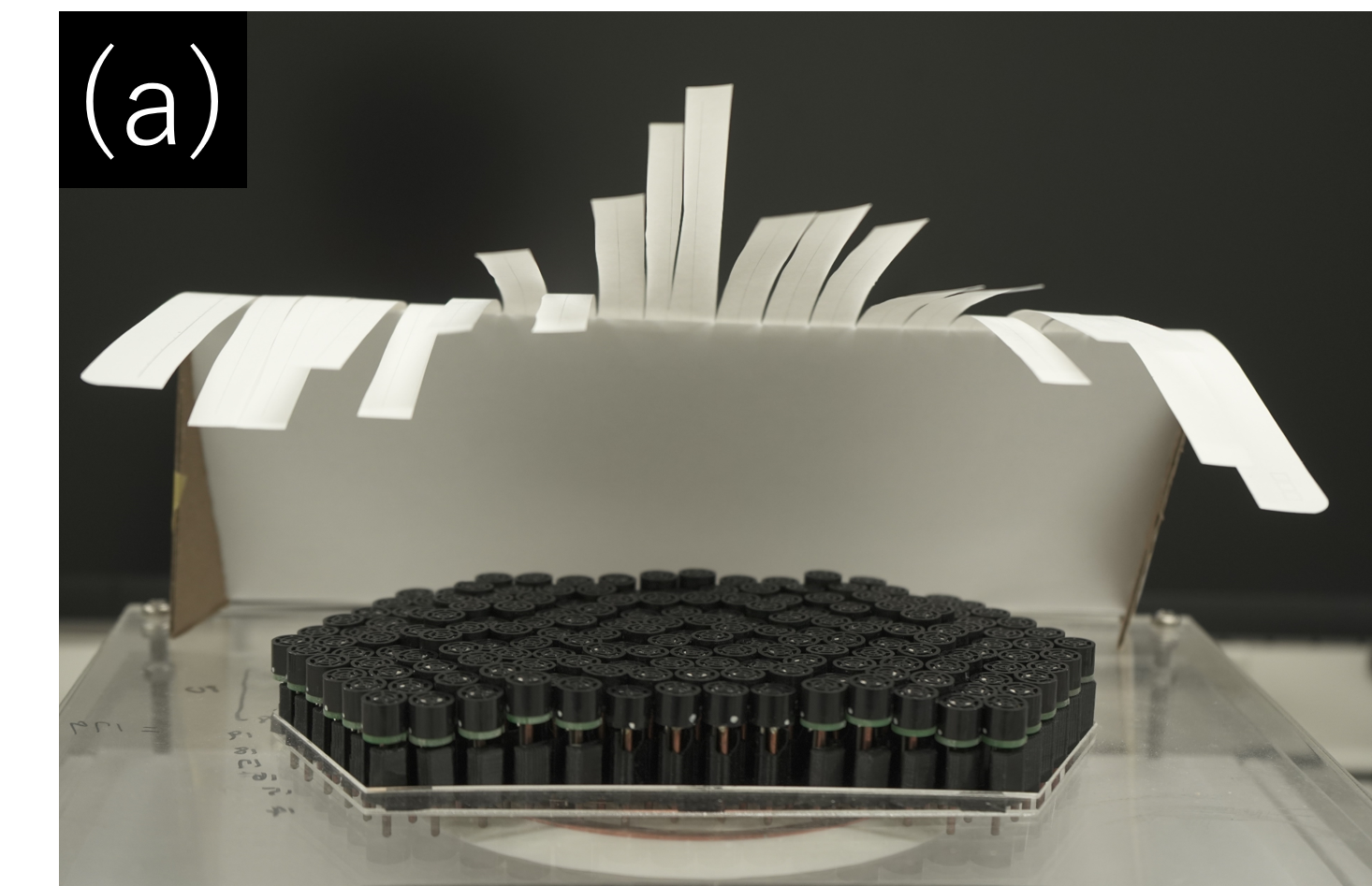}
    \label{subfig:applicationa}
  \end{minipage}
  \begin{minipage}[b]{0.8\columnwidth}
    \includegraphics[width=\columnwidth]{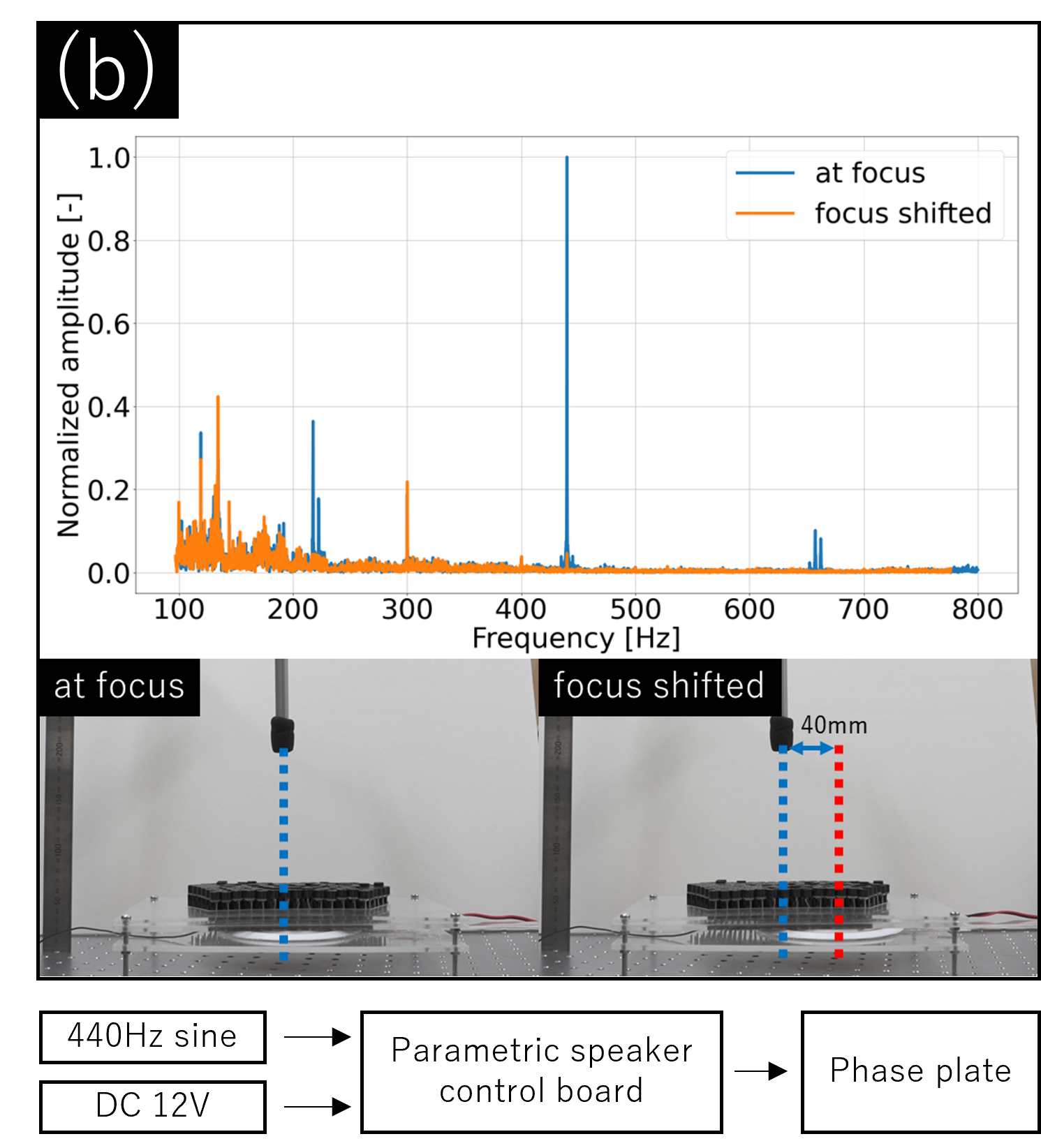}
    \label{subfig:applicationb}
  \end{minipage}
\caption{Practical applications of the manually reconfigurable phased array system. (a) Demonstration of haptic feedback: A spherical phase plate (radius 150 mm) creates a focused acoustic field, causing visible displacement of a paper strip with a 10 mm wide slit placed above the array. (b) Audio spot technology application: The graph shows how adjusting the phase plate position controls the location of the audio spot. The experiment used a 440 Hz sine wave input from a parametric speaker test kit, with fixed positions for the microphone and array.}
\label{fig:application}
\end{figure}
We experimentally demonstrated the pressure field using a Laser Doppler Vibrometer (LDV, Polytec-500-3D-HV-Xtra) to measure the sound pressure distribution in the vertical plane relative to the transducer array \cite{Malkin2014}. For this experiment, we used a phase plate in Figure~\ref{fig:phasePlate}d designed to generate a Bessel beam.
The experimental set up included a phase plate in Figure~\ref{fig:phasePlate}d with a retroreflective sheet placed behind the array to enhance LDV measurements. We took measurements at three points on the phase plate: at the center (0 m) and at ±0.0357 m along the x-axis relative to the array's center.
Figure~\ref{LDV} presents both the simulated and measured results. The simulation results were obtained using the equation provided in the supplementary material, summed along the y-axis and normalized against the maximum pressure. As demonstrated in the figure, the simulated and measured sound pressure distributions closely align, thereby validating our theoretical model. However, some discrepancies were observed in the characteristics of the side lobes when compared to the simulation results. The presence of noise observed in the experiment, particularly in the range of x-axis 0.05 to 0.1, as seen in Figures 3a and 3b, is attributed to an experimental artifact caused by air bubbles within the retroreflective sheet. Despite these differences, the overall agreement between the simulation and experiment confirms the effectiveness of our manually reconfigurable phased array in generating predictable acoustic fields, particularly in creating a Bessel beam.

We investigated two practical applications of our manually reconfigurable phased array system: haptic feedback and audio spot technology.
As illustrated in Figure~\ref{fig:application}a, this phase plate in Fig.\ref{fig:phasePlate}d was placed on the underside of the transducer array. We placed a paper with a 10 mm wide slit on the upper surface of the array. When activated, the acoustic radiation force at the focal point visibly pushed the paper upwards, demonstrating the system's potential for creating localized, tactile sensations in mid-air. This result suggests the feasibility of using our system for non-contact haptic feedback applications.
We also explored the system's capability for audio spot technology. In this experiment, we kept the positions of the microphone and the array fixed while examining changes in parametric sound as we moved the phase plate. We used a 440 Hz sine wave as the input sound source, generated by the control board from TriState's "Parametric Speaker (Single-directional Speaker) Test Kit." Figure~\ref{fig:application}b shows the results of this experiment. By adjusting the position of the phase plate, we were able to control the location of the audio spot effectively. This demonstrates the system's potential for creating localized sound fields, which could be valuable in applications such as personalized audio zones or directional audio systems.

In conclusion, our manually reconfigurable phased array system, consisting of a phase plate and retractable transducers, offers an approach to acoustic wavefront manipulation without complex FPGA boards. This design, with its adaptability and customization potential, has applications in various acoustic research fields. Moreover, the system extends beyond traditional research applications into educational settings. By allowing hands-on manipulation of acoustic fields, it makes the concept of acoustic holograms more tangible and accessible to students. This practical demonstration enhances the understanding of complex acoustic phenomena, bridging the gap between theoretical acoustics and its physical manifestation.

\begin{acknowledgments}
S.M. was sponsored by the Mitou program of the Information-Technology Promotion Agency, Japan, under the project management of Professor Masahiko Inami. We extend our thanks to Pixie Dust Technologies, Inc. for providing access to the Laser Doppler Vibrometer (LDV) used in our experiments.
\end{acknowledgments}

\section*{Data Availability Statement}
The data that supports the findings of this study are openly available in Zenodo at [DOI] and its supplementary material. 
\bibliography{references}

\end{document}


\begin{flushleft}

\title{Spatial Sound Modulation through Manual Reconfiguration with Phased Plate}
\author{Soma Mochizuki}
\affiliation{\footnotesize{School of Informatics, College of Media Arts, Science and Technology, University of Tsukuba, Kasuga Campus Kasuga 1-2, Tsukuba, 305-8550, Ibaraki, Japan}}
\author{Yoichi Ochiai}
\affiliation{\footnotesize{Institute of Library, Information and Media Science, University of Tsukuba, Kasuga Campus Kasuga 1-2, Tsukuba, 305-8550, Ibaraki, Japan}}
\affiliation{\footnotesize{R\&D Center for Digital Nature, University of Tsukuba, Kasuga Campus Kasuga 1-2, Tsukuba, 305-8550, Ibaraki, Japan}}
\affiliation{\footnotesize{Pixie Dust Technologies, Inc, Tokyo, 104-0028, Japan}}
\author{Tatsuki Fushimi}
\affiliation{\footnotesize{Institute of Library, Information and Media Science, University of Tsukuba, Kasuga Campus Kasuga 1-2, Tsukuba, 305-8550, Ibaraki, Japan}}
\affiliation{\footnotesize{R\&D Center for Digital Nature, University of Tsukuba, Kasuga Campus Kasuga 1-2, Tsukuba, 305-8550, Ibaraki, Japan}}

\date{\today}
\maketitle

\section{Simulation}
The pressure at a specific point in the acoustic field is calculated using the following finite sum equation:
\begin{equation}
    p = \sum_{n=1}^{N} \left( \frac{p_0}{R_{n}} \cdot D_{n} \cdot \exp\left(j \cdot \left(k \cdot R_{n} + \phi_{n}\right)\right) \right).
\end{equation}
The sum is taken over all $N$ transducers. For each transducer $n$, $p_0$ is the transducer amplitude at 1 m, $R_{n}$ is the distance from the $n$-th transducer to the point in the 3D grid, and $D_{n}$ is the directivity function for the $n$-th transducer. The phase of the $n$-th transducer is determined and incorporated into the calculation based on the position of the phase plate. The state of the transducer can be inactive, in-phase, or anti-phase.
The directivity is calculated using the following equation:
\begin{equation}
D_{n} = 
\begin{cases}
\frac{2 \cdot J_1\left( k \cdot a \cdot \sin(\theta) \right)}{k \cdot a \cdot \sin\left(\theta\right)}, & \theta \neq 0\\
1, & \theta = 0.
\end{cases}
\end{equation}

The expression $\exp\left(j \cdot \left(k \cdot R_{n} + \phi_{n}\right)\right)$ is the complex exponential function that incorporates the phase change due to the distance \( R_{n} \) and the additional phase contribution \( \phi_{n} \) from the \( n \)-th transducer. Here, \( j \) represents the imaginary unit, and \( k \) is the wavenumber, calculated as \( k = \frac{2\pi}{\lambda} \).

\section{Hardware}
\subsection{Build instructions}
\subsubsection{Transducer Module}
Supplementary Figure1 shows the components required to create the transducer module.
\begin{figure}[ht]
    \centering
    \includegraphics[width=0.8\textwidth]{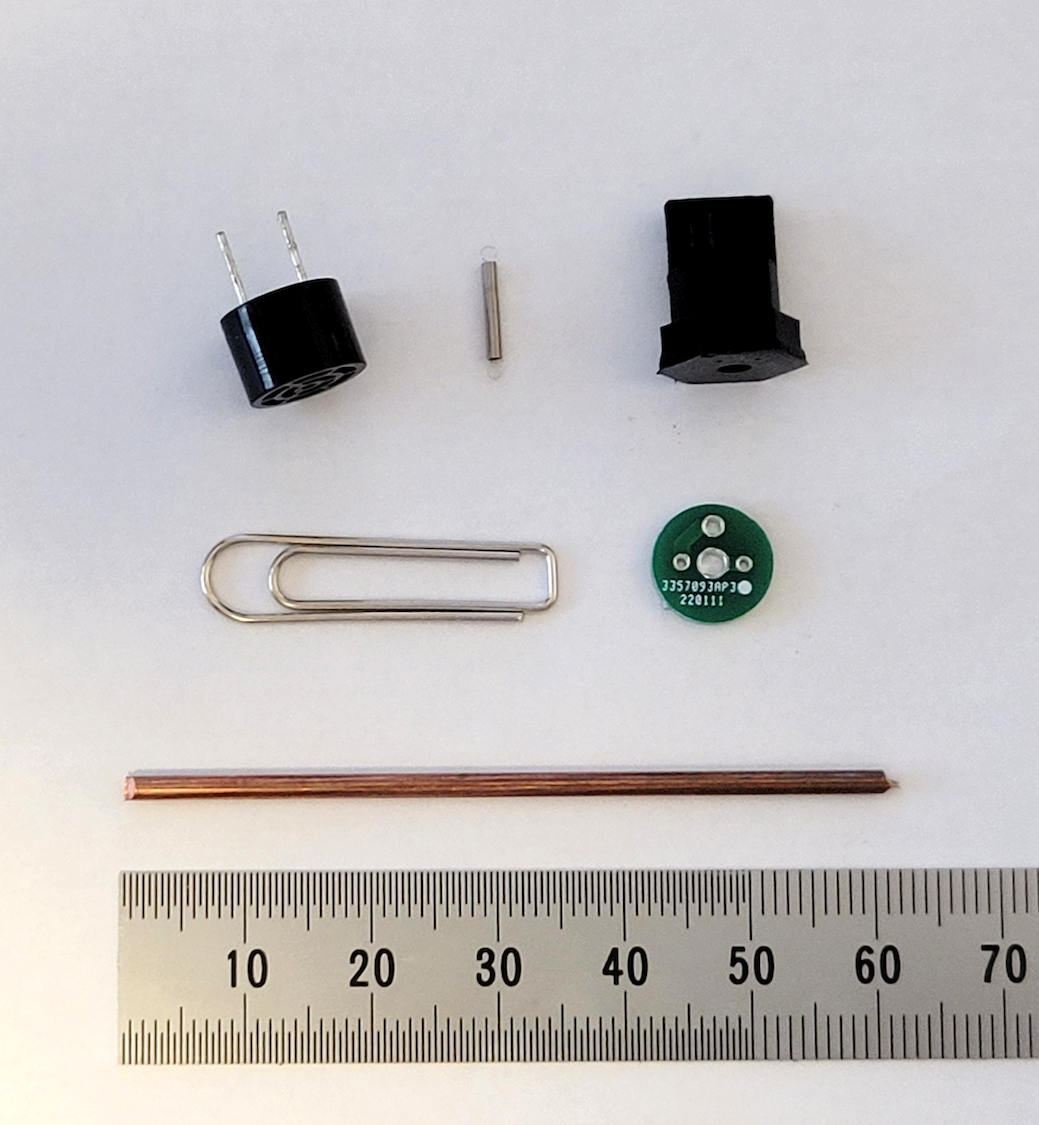}
    \caption{Components for the transducer assembly}
\end{figure}
\begin{figure}[ht]
    \centering
    \includegraphics[width=0.8\textwidth]{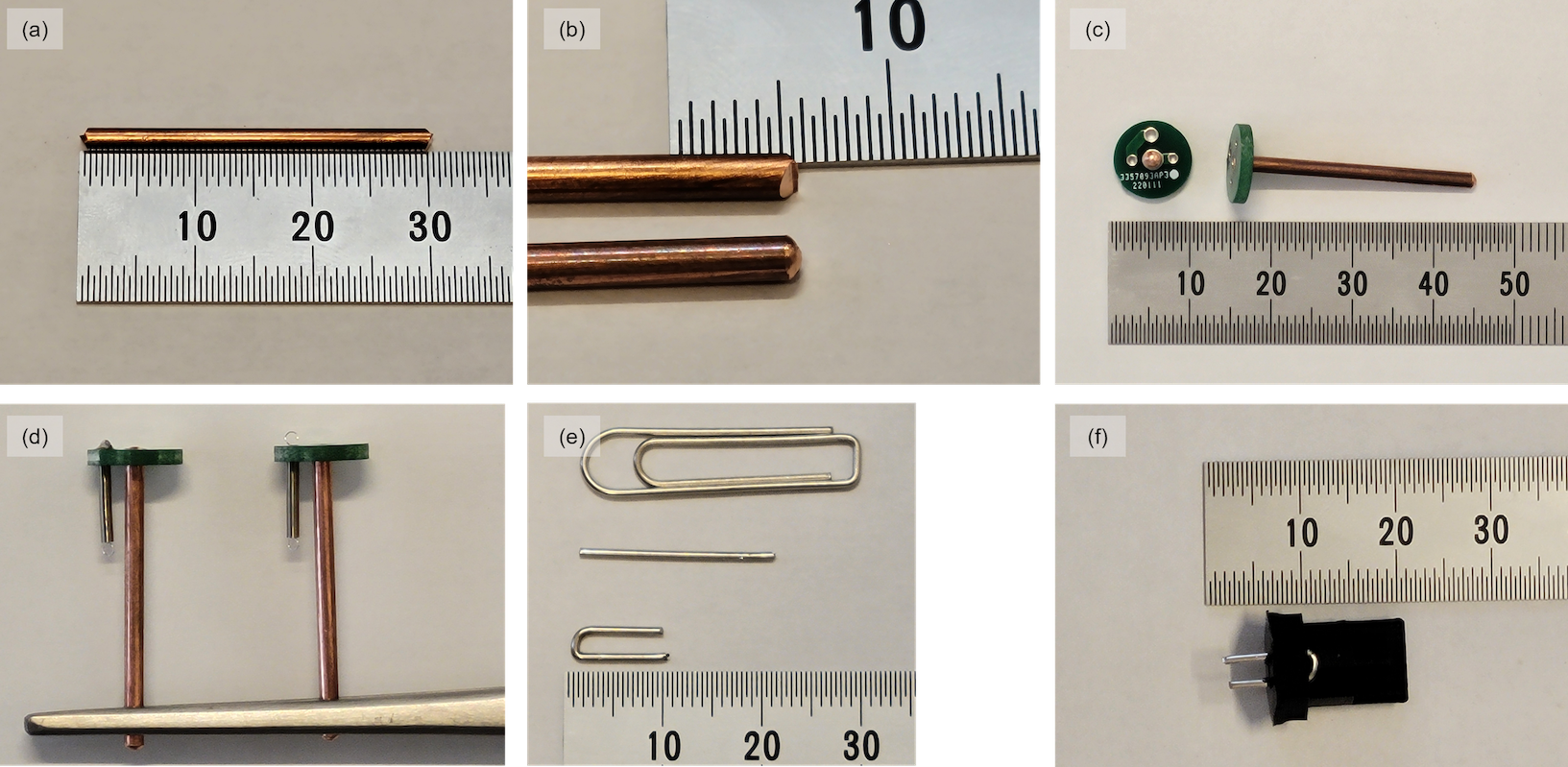}
    \caption{Building steps of transducer assembly}
\end{figure}
\begin{figure}[ht]
    \centering
    \includegraphics[width=0.8\textwidth]{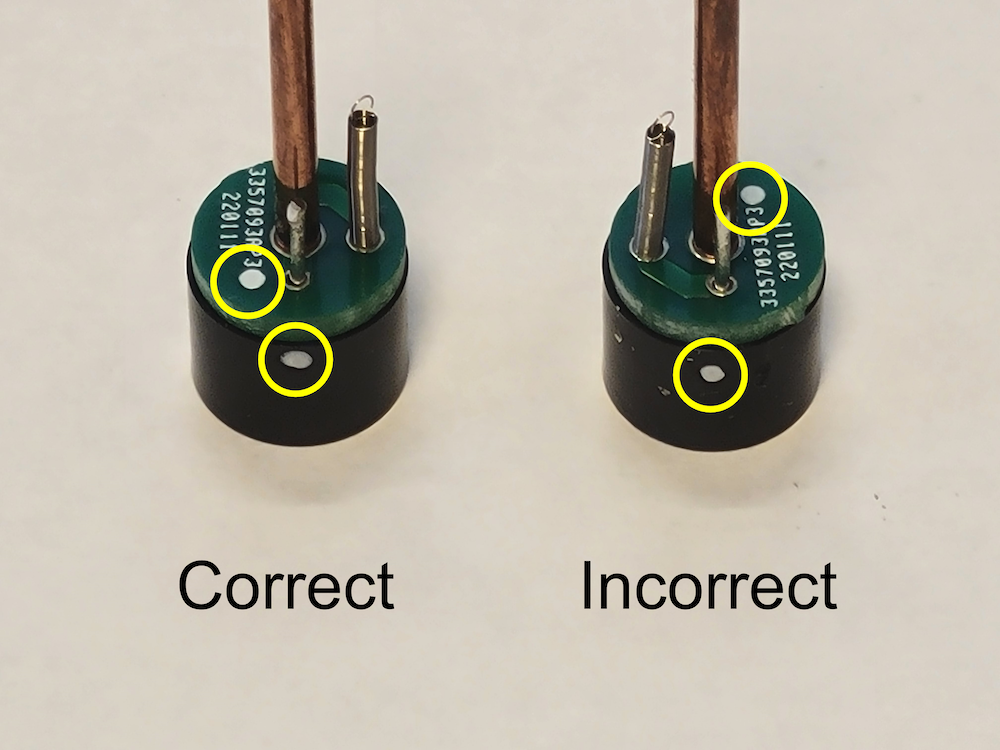}
    \caption{Correct Orientation of the Transducers and PCB}
\end{figure}
\begin{enumerate}
    \renewcommand{\labelenumi}{(\arabic{enumi})}
    \item Supplementary Figure2 a. Cut copper rods to a length of 30mm, and create 179 copper rods of this length. (The number of copper rods can be adjusted according to the desired number of transducers for the transducer array.)
    \item Supplementary Figure2 b. Process one end face of each copper rod to be hemispherical using a file. This is a necessary step to prevent the copper rods from interfering and damaging with the operation of the phase plate when it is moved.
    \item Supplementary Figure2 c. Insert the non-hemispherical end face of the copper rods into the through-holes at the center of the PCB. If the copper rods cannot be secured to the PCB, solder them in place.
    \item Supplementary Figure2 d. Solder springs to the PCB with the attached copper rods. Push the hook part at the end of the spring through the through-hole on the same side as the inserted copper rod and solder it in place. Be mindful of the polarity of the through-hole for soldering the spring and the PCB surface. Additionally, only solder the end of the spring, as soldering the entire spring may cause damage or changes in its properties.
    \item Supplementary Figure3. The ultrasonic transducer is soldered onto the substrate. It is crucial to heed the polarity of the transducer during this process, utilizing the white circle on the substrate as a reference point.
    \item Supplementary Figure2 e. Straighten and cut clips to a length of about 2cm to create 179 metal wires.
    \item Supplementary Figure2 e. Bend the metal wires with pliers to create hooks in the shape shown in the diagram.
    \item Supplementary Figure2 f. Heat press the hooks into the 3D printed parts with a soldering iron.
    \item While inserting the copper rods into the holes at the center of the 3D printed parts, solder the ends of the springs to the hooks. Assemble all components in the same manner.
\end{enumerate}
\subsection{Transducer Array}
\begin{enumerate}
    \renewcommand{\labelenumi}{(\arabic{enumi})}

    \item Attach the completed parts together using an acrylic adhesive or similar solvent to create the transducer array. Adhere the hexagonal parts of the 3D printed components in the densest arrangement possible. During this process, try to make the array as flat as possible.
    \item Using a laser cutter, process the acrylic plates as shown in the diagram. Create three plates in total.
    
    \item Sandwich the array from top to bottom with a pair of plates with holes in the center and secure it.
    \item Combine the last acrylic plate according to the holes in the four corners. At this time, ensure that the probe is about 3mm away from the acrylic on the bottom.
\end{enumerate}

\end{flushleft}